# Pulse width controlled resistivity switching at room temperature in $Bi_{0.8}Sr_{0.2}MnO_3$

A. Rebello and R. Mahendiran[*]

Department of Physics and NUS Nanoscience & Nanotechnology Initiative (NUSNNI), Faculty of Science, National University of Singapore, 2 Science Drive 3, Singapore -117542

**Abstract**

We report pulsed as well as direct current/voltage induced electroresistance in $Bi_{0.8}Sr_{0.2}MnO_3$ at room temperature. It is shown that bi-level and multi-level resistivity switching can be induced by a sequence of pulses of varying pulse width at fixed voltage amplitude. Resistivity increases abruptly ($\approx$ 55 % at 300 K) upon reducing pulse width from 100 ms to 25 ms for a fixed electric field ($E$ = 2 V/cm$^2$) of 200 ms pulse period. The resistivity switching is accompanied by a periodic change in temperature which alone can not explain the magnitude of the resistivity change.

---

[*] Corresponding author – phyrm@nus.edu.sg

Materials that show switching between high ("OFF") and low ("ON") resistance states upon the change in current/voltage amplitude are currently attracting attention due to their potential applications as resistive random access memory (ReRAM) elements in next generation nonvolatile memories (NVM).[1] Resistivity switching in metal-insulator-metal (MIM) layered structure has been reported recently in a number of binary and tertiary oxides such as NiO,[2] TiO2,[3] Cr- doped $SrZrO_3$,[4] $La_{2-x}Sr_xNiO_4$,[5] $La_{0.33}Sr_{0.67}FeO_3$,[6] $Pr_{0.7}Ca_{0.3}MnO_3$,[7] $LuFe_2O_4$,[8] $Na_{0.5-\delta}CoO_2$ [9] etc. The exact origin of the resistivity switching in materials belonging to such a wide range of oxides is still elusive. The Mn-based oxides (manganites) offer certain advantages over other oxides since they exhibit a variety of electronic and magnetic ground states as a function of temperature and chemical doping, and their resistivity can be tuned both by electric current as well as magnetic field. Interestingly, it was also reported that current-induced electroresistance can lead to a change in magnetization in some manganites such as $Pr_{1-x}Ca_xMnO_3(x \approx 0.33)$.[10,11]

For manganites alone, several mechanisms such as modification of the Schottky barrier height,[12] trap controlled space-charge limited current,[13] electric field-induced melting of charge ordering,[7] excitation of charge density waves,[14] formation of conductive filamentary paths,[10] small-polaron-hopping,[15] polaron order-disorder transition at the metal-oxide interface,[16] Mott transition at the interface,[17] creation of defects and migration of oxygen ions,[18] have been proposed to explain the current or voltage-induced electroresistance. The electrical resistivity switching induced by pulsed voltage and current received more attraction following the initial report of polarity dependent resistance change in $Pr_{0.7}Ca_{0.3}MnO_3$ sandwiched between two metal electrodes, high resistance state for positive voltage pulses and low resistance for negative voltage pulses at room temperature.[19] The magnitude of electroresistance depends not only on the polarity but also on the amplitude and number of pulses. However, the influence of pulse width on the electroresistance effect has been studied in less detail. Odagawa *et al.*[13] found that the electroresistance is independent of pulse width from *dc* to 150 ns in $Pr_{0.7}Ca_{0.3}MnO_3$ film.

In this work, we report *dc* and pulsed *I-V* characteristics and resistivity switching with a sequence of pulses of different pulse width along with response of sample's surface temperature during the current and voltage pulsing in $Bi_{0.8}Sr_{0.2}MnO_3$ (BSMO). Though $Bi_{1-x}Sr_xMnO_3$ series has been studied from the view points of structural and magnetic properties,[20] non linear electrical transport in these compounds has not been reported so far. Upon increasing the Sr content, the ground state of $Bi_{1-x}Sr_xMnO_3$ transforms from a ferromagnetic insulator ($T_C$ = 105 K for x = 0) to an antiferromagnetic and charge ordered insulator for x ≥ 0.25. The main objectives of this report are to investigate the origin of the nonlinearity and hysteresis seen in the *J-E* characteristics and also to explore the possibility of pulsed current induced resistivity switching in BSMO. The titled compound does not suffer from the problem of mesoscopic phase separation which is commonly encountered in other manganites.

We show that BSMO exhibits a direct current (*dc*) induced electroresistance effect and simultaneous measurements of *dc I-V* characteristics and surface temperature of the sample reveal that the strong nonlinearity and hysteresis observed in these two compounds at higher current density is caused by the Joule heating. However, in the voltage regime of negligible Joule heating, the resistance can be reversibly switched between a high and a low resistance states by controlling the pulse width of the voltage excitation even if the voltage amplitude is fixed. We also demonstrate multi-level resistivity switching controlled by the width of pulsed current and retention behavior of memory states in these materials.

The four probe *dc* and pulsed *I-V* characteristics of a bar shaped polycrystalline BSMO sample (10 mm x 2.5 mm x 3 mm) at different temperatures were measured using source-measure units (Keithley 2400 and Yokagawa GS610) interfaced to the Physical Property Measurement System (PPMS, Quantum Design Inc, USA). The electrical contacts were made with Ag-In alloy or Ag paste and the results were found to be identical. The sample was glued to a thin mica substrate with GE-7031 varnish which was anchored to the standard PPMS brass sample puck. To monitor the temperature of the sample a Pt-thermometer (PT100) of size 3x2x1.2 mm was glued to the top surface of the sample with a good thermal conductive grease (Apiezon-N grease) and a small quantity of GE-7031 varnish. The Pt resistor was positioned between two voltage probes that were separated by 6.5 mm. The four probe resistance of the Pt thermometer was monitored by measuring the voltage across the Pt-resistor using a Keithley 2182A nanovoltmeter while supplying a constant current of 10 µA with Keithley 6221 dc and ac current source. The cernox sensor attached to the sample puck measured the base temperature by PPMS. It is to be pointed out that the base temperature recorded by the PPMS did not show any variation (remains at the stable value of 300 K) with current/voltage sweep and this leads to the general misconception of a trivial role played by joule heating in the sample. On the contrary, since the top surface of the sample is in very good thermal contact with the PT100 thermometer in our experimental set up, we could measure a significant change in the temperature of the sample during the current/voltage sweeps. Although a small temperature difference between thermometer and the sample is possible, the difference will be one tenth of kelvin and hence negligible. All the pulsed measurements were performed using a Yokogawa GS610 source-measure unit.

Figure 1(a) shows temperature dependence of the resistivity ($\rho$) of the sample for two different *dc* current strengths, $I = 100$ µA and 20 mA on the left scale. The x-axis shows the temperature ($T$) of the sample recorded by the PPMS which we call as the "base temperature" and the right scale shows "surface temperature" of the sample measured by the platinum resistance sensor attached to the top surface of the sample. The inset shows the experimental configuration. While the resistivity curves for both values of the current matches above 300 K, they clearly bifurcate below 300 K. The $\rho$ ( $I = 100$ µA) continues to increase with lowering temperature until $T = 100$ K whereas $\rho$ ( $I = 20$ mA) is lower in magnitude by few orders and shows an apparent saturation behavior below 180 K. Remarkably, $\rho$ decreases as much as 99 % at 50 K as the current increases from 100 µA to 20 mA. However, we find that the current-induced electroresistance is accompanied by a large increase in the surface temperature of the sample. When the base temperature of the sample, read by the thermometer in the PPMS is 50 K, the surface temperature of the sample is as high as 150 K for 20 mA suggesting a significant Joule heating of the sample as indicated on the right scale. Since the sample is semiconducting, Joule heating can contribute to the lowering of resistance in addition to other non-thermal mechanisms. This type of

current –induced electroresistance was reported earlier in $Nd_{1-x}Ca_xMnO_3$ (x = 0.3, 0.5) but possible Joule heating was not considered by the authors.[21] When the resistivity is plotted versus the surface temperature of the sample, data for both the current amplitudes match above 100 K (not shown here).

Figure 2(a) shows the electric field (*E*) versus current density (*J*) behavior at room temperature while sweeping the current density in both *dc* (open circle) and pulsed current (line) modes. For pulsed current sweep we used pulses of a long period (PD = 1 s) and short pulse width (PW = 25 ms), increasing the amplitude of the pulse in step of 0.02 mA. The response of the surface temperature of the sample during the pulsed (open square) and *dc* (line) current density sweep is shown on the right scale. Non linear *E-J* behavior is seen for both the *dc* and pulsed currents. While the electric field shows a tendency to saturate at *dc* current density higher than 0.2 A/cm$^2$, it continues to increase without saturation in the pulsed current sweep. However, the increase in the surface temperature of the sample is only 0.5 K in the pulsed mode compared to an increase of 25 K in the *dc* mode. The Joule-heating assisted enhancement in the *dc* electroresistance was also reported in metallic $La_{0.8}Ca_{0.2}MnO_3$ film and also in the phase separated Cr- doped $Nd_{0.5}Ca_{0.5}MnO_3$ though effect of pulsed current was not investigated in those studies.[22]

In the *dc* voltage sweep mode, (see Fig. 2(b)), *J* varies gradually for *E* below 1 V/cm but a rapid increase occurs above *E* = 1.5 V/cm. Upon reducing the *E* from the maximum value, *J* shows a pronounced hysteresis at high electric field but merges with the initial curve close to the origin. However, the rapid increase in the current density at higher electric field strength in the forward sweep is invariably accompanied by a rise in temperature (~ 25 K). The high field non linearity is eliminated and heating effect is strongly reduced in the pulsed *E*- sweep mode. Tokunaga *et al.*[23] observed much sharper *dc* voltage jump at a threshold *dc* current in the phase separated manganite La-Pr-CaMnO$_3$ and attributed it to the local Joule heating effect. Interestingly, an anomalous behavior of oscillation in the current was also found if the voltage was maintained at a constant value and this was attributed to the closing and opening of metallic conducting channels.

The striking differences between the *dc* and pulsed *J-E* characteristics motivated us to investigate the response of electrical resistivity of the BSMO to a sequence of pulses with varying pulse width (PW) but with fixed voltage amplitude. We have chosen, $E$ = 2 V/cm$^2$ to avoid excessive Joule heating. Fig. 3(a) shows the change in $\rho$ (left scale) and Tsurface (right scale) in response to four pulse trains of 2 V/cm$^2$ amplitude with a fixed pulse period (PD = 1 s) but two different pulse width (PW = 100 ms and 25 ms). Each pulse train consists of 1000 pulses of PW = 25 ms followed by another 1000 pulses of PW = 0.1 s and the start of the new pulse train is marked by arrows in the figure. After the initial drop of resistance at the beginning of the pulse train, the resistance abruptly jumps up by 6 % when the PW is changed from 100 ms to 25 ms and remains nearly unchanged until another pulse of a longer pulse width (PW = 100 ms) sets it to a low resistance state. The resistance can be set to a higher value again by the application of the shorter pulse width (PW = 25 ms). In accordance with the resistivity switching, temperature of the sample also changes periodically though only by ~ 0.5 K. If the period of the pulse is lowered to 200 ms (see fig. 3 (b)) change in the pulse width from 25 ms to 100 ms results in decrease of $\rho$ by 55 % but *T* also increases by 8 K. We have calculated the change in $\rho$ as *T* increases from 300 K to 308 K from $\rho(T)$ vs. $T$ curve (fig. 1) and it turns out that $\rho$ decreases by about 23 % which is smaller than found with the pulse width –induced switching. This suggests that electroresistance can occur partially due to the Joule heating but other non-thermal mechanisms may also play a significant role.

Interestingly, we can also induce tri-level resistivity switching by introducing three different pulse widths as shown in fig. 3(c). A pulse train of PW = 0.1 s decreases $\rho$ by 40 % whereas decreasing PW to 50 ms leads to an increase of $\rho$ by 45 %. Further decreasing PW to 25 ms, $\rho$ increases by 14 %. The temperature of the sample also changes accordingly. The repetition of pulse sequences drives the sample to the respective resistive states and is completely reproducible. All these results were also reproducible if the measurements were done at ambient temperature and pressure outside the PPMS set up (data not shown here).

At present, the exact origin of the pulse width controlled resistance switching is not clearly understood. An obvious possibility is the local Joule heating. The longer the pulse width, the input power (P = $(V^2/R)t_{pw}$, where $t_{pw}$ is the pulse width (pulse duration) and R is the resistance of the sample) gets enhanced and causes the decrease in the resistance as a result of the Joule heating for a period $t_{pw}$. When the input power is zero between the 'ON' states of consecutive pulses, the sample will cool according to the Newton's law of cooling and this will lead to the increase in the resistance. When the sample is subjected to a sequence of pulses, the sample temperature reaches a thermal equilibrium after a certain number of pulses. This leads to a sharp increase of resistance initially and then a saturation after a certain number of pulses as can be seen clearly in Figs. 2(b) and 2(c). Similar behavior of resistance change with various voltage amplitudes but for a fixed pulse width and period was observed in the phase separated Cr-doped Nd-Ca-MnO$_3$ by Song *et al.*[24] In our case, when the pulse width is periodically changed for a given period, the sample temperature also changes periodically and leads to a periodic changes in the resistance. The change in the resistance is large if the period is small as suggested by the comparison of figure 3(a) and 3 (b). However, as mentioned earlier, the percentage change in the resistance we obtained from the pulsed mode is larger than the change in the resistance obtained by considering the Joule heating alone from the $\rho$-$T$ curve and therefore other non thermal mechanisms have to be also considered. Recently, it has been shown that current injection from the tip of a scanning tunneling microscope can induce local melting of charge ordering and hence the resistivity switching is of non-thermal origin in Pr$_{0.67}$Ca$_{0.33}$MnO$_3$.[25] Interestingly, three-orders of magnitude decrease in resistivity was also reported in Pr$_{0.7}$Ca$_{0.3}$MnO$_3$ with excitation of a specific phonon mode (17 THz) related to large amplitude Mn-O distortion rather than to Joule heating.[26] According to a recent model proposed by Quintero *et al.*[27] there is a variation of the concentration of oxygen vacancies near the electrode-sample interface ("interfacial domains"). Under the action of alternative pulsing, oxygen vacancies move back and forth between the bulk central region ("bulk") and interfacial regions, thus effectively doping the central region. Thus, the pulsed voltage switching modifies the carrier concentration and it could probably explain our observation. A systematic study of pulse width dependence of resistivity switching from milli to pico second range may be helpful to clearly understand the origin of the resistivity switching in these oxides.

R. M acknowledges Ministry of Education (Singapore) for supporting this work through the grant  ARF/Tier-1: R-144-000-167-112.

**Figure captions:**

Fig. 1 Temperature dependence of the resistivity of $Bi_{0.8}Sr_{0.2}MnO_3$ for *dc* current values $I$ = 100μA and 20 mA. The *x*-axis shows the base temperature of the sample measured by the cernox sensor installed in the cryostat close to the sample holder. The *y*-axis on the right scale shows the temperature measured by the Pt-sensor glued on the top surface of the sample.

Fig. 2 The *dc* and pulsed (a) Current density ($J$) dependence of the *dc* electric field($E$) and (b) Electric field ($E$) dependence of the current density ($J$) at $T$ = 300 K.

Fig. 3 (a) Resistivity switching at $T$ = 300 K triggered by changes in pulse widths under different periods, (a) PD = 1 s, (b) PD = 200 ms and (c) tri-level resistivity switching for three different pulse widths for PD = 200 ms. The numbers in the graphs indicate pulse widths.

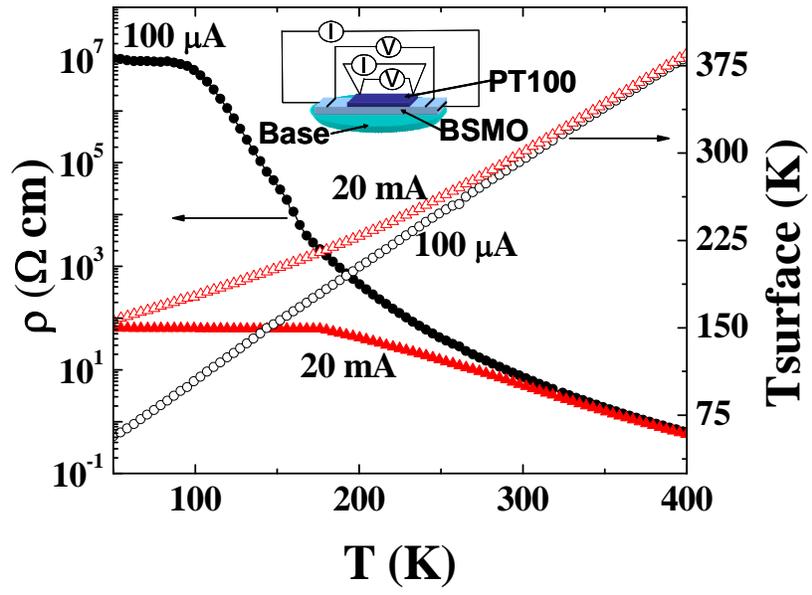

Fig. 1

A. Rebello *et al*.

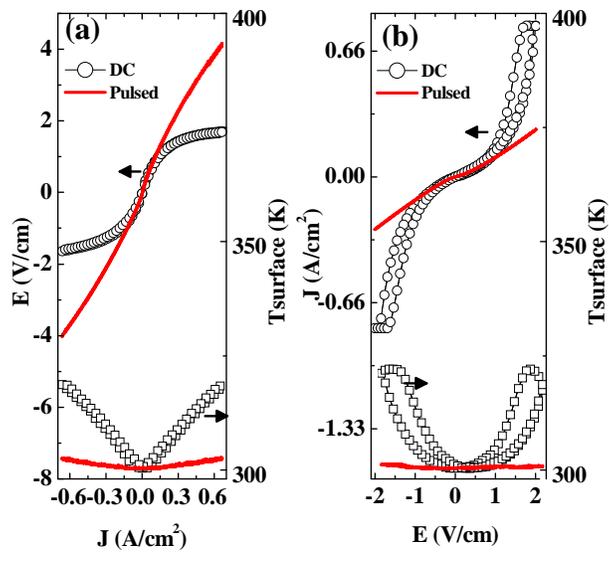

Fig. 2

A. Rebello *et al*.

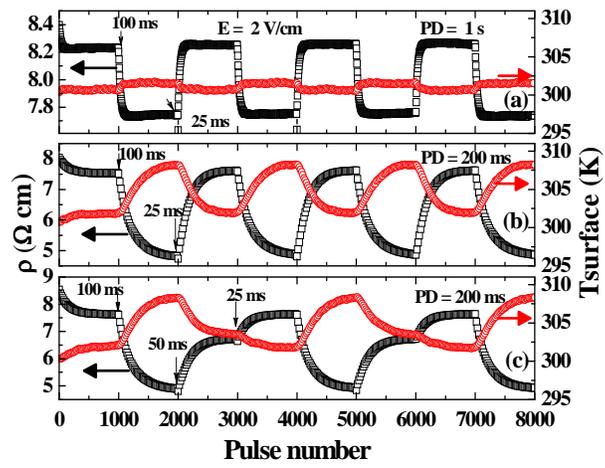

Fig. 3
A. Rebello *et al*.